\documentclass{article}
\usepackage{braket,bm,latexsym,amsmath,amssymb,amsfonts,fancyhdr,color,graphicx,multirow,slashed,cite}
\usepackage[a4paper,bottom=3cm,top=2.5cm,head=0mm,width=17cm,dvipdfm]{geometry}
\usepackage[usenames,dvipsnames,svgnames,table]{xcolor}
\usepackage[colorlinks=true,
            linkcolor=blue,
            urlcolor=blue,
            citecolor=green,          
						bookmarks=true,
						bookmarksnumbered=true,
						breaklinks=true,
						pdfpagemode=Fullscreen,
						pdfstartview=FitBH]{hyperref}
\usepackage[dotinlabels]{titletoc}
\usepackage{titlesec}
\usepackage{authblk,ulem}

\numberwithin{equation}{section}
\allowdisplaybreaks[4]

\titlelabel{\thetitle.\quad \hspace{-0.8em}}
\titlecontents{section}
              [1.5em]
              {\vspace{4mm} \large \bf}
              {\contentslabel{1em}}
              {\hspace*{-1em}}
              {\titlerule*[.5pc]{.}\contentspage}
\titlecontents{subsection}
              [3.5em]
              {\vspace{2mm}}
              {\contentslabel{1.8em}}
              {\hspace*{.3em}}
              {\titlerule*[.5pc]{.}\contentspage}
\titlecontents{subsubsection}
              [5.5em]
              {\vspace{2mm}}
              {\contentslabel{2.5em}}
              {\hspace*{.3em}}
              {\titlerule*[.5pc]{.}\contentspage}

\newcommand{\titledef}{F\'eeton ($B-L$ Gauge Boson) Dark Matter Testable in Future Direct Detection Experiments} 

\hypersetup{ pdfauthor = {Shao-Feng Ge},
	     pdftitle = {\titledef}, 
	     pdfsubject = {}, 
             pdfkeywords = {}, 
	     pdfcreator = {LaTeX with hyperref package},
	     pdfproducer = {dvips + ps2pdf} }

\definecolor{gesfblack}{rgb}{0,0,0}

\definecolor{gesfblue}{rgb}{0.08,0.42,0.76}

\definecolor{gesfgreen}{rgb}{0,1,0}

\definecolor{gesfgrey}{rgb}{0.5,0.5,0.5}

\definecolor{gesflanse}{rgb}{0.00,0.50,0.50}

\definecolor{gesfpurple}{rgb}{0.47,0.19,0.42}

\definecolor{gesfred}{rgb}{1,0,0}

\definecolor{gesfwhite}{rgb}{1,1,1}

\definecolor{gesfyellow}{rgb}{0.7,0.4,0.3}

\newcommand{\geqn}[1]{\hypersetup{linkcolor=blue}Eq.\,(\ref{#1})\hypersetup{linkcolor=blue}}
\newcommand{\gfig}[1]{{\hypersetup{linkcolor=violet}Fig.\,\ref{#1}\hypersetup{linkcolor=blue}}}

\definecolor{Orange}{cmyk}{0,0.61,0.87,0}
\definecolor{JungleGreen}{cmyk}{0.99,0,0.52,0}
\definecolor{OliveGreen}{cmyk}{0.64,0,0.95,0.40}
\definecolor{Brown}{cmyk}{0,0.81,1,0.60}
\definecolor{RoyalBlue}{cmyk}{0.71,0.53,0,0.12}
\definecolor{Gray}{cmyk}{0,0,0,0.40}
\definecolor{LightPink}{cmyk}{0.0,0.25,0,0}
\definecolor{LLightPink}{cmyk}{0.0,0.10,0,0}
\definecolor{LightBlue}{cmyk}{0.25,0,0,0}
\definecolor{LightGray}{cmyk}{0,0,0,0.2}

\graphicspath{{figs/}}

\usepackage{subfigure}
\newcommand{\bee}{\begin{equation}}
\newcommand{\ene}{\end{equation}}
\newcommand{\bea}{\begin{eqnarray}}
\newcommand{\ena}{\end{eqnarray}}

\begin{document}
\fontsize{12pt}{14pt}\selectfont

\title{
       \Large \bf \titledef} 
\author[4,1,2]{{\large Yu Cheng} \footnote{\href{mailto:chengyu@sjtu.edu.cn}{chengyu@sjtu.edu.cn}}}
\affil[1]{Tsung-Dao Lee Institute \& School of Physics and Astronomy, Shanghai Jiao Tong University, Shanghai 200240, China}
\affil[2]{Key Laboratory for Particle Astrophysics and Cosmology (MOE) \& Shanghai Key Laboratory for Particle Physics and Cosmology, Shanghai Jiao Tong University, Shanghai 200240, China}
\author[1,2]{{\large Jie Sheng} \footnote{\href{mailto:shengjie04@sjtu.edu.cn}{shengjie04@sjtu.edu.cn}}}
\author[3,1,2]{{\large Tsutomu T. Yanagida} \footnote{\href{mailto:tsutomu.tyanagida@sjtu.edu.cn}{tsutomu.tyanagida@sjtu.edu.cn}}}
\affil[3]{Kavli IPMU (WPI), UTIAS, University of Tokyo, Kashiwa, 277-8583, Japan}
\affil[4]{Department of Physics, KAIST, Daejeon, 34141, Korean}
\date{}

\maketitle

\vspace{-2mm}
\begin{abstract}
\fontsize{10pt}{12pt}\selectfont

In this paper, we revisit the f\'eeton (gauge boson of $U(1)_{B-L}$ symmetry) dark matter scenario, and first point out the 
$U(1)$ gauge symmetry can be a linear combination of the $B-L$  and the SM
hypercharge gauge symmetries. With the redefinition of $B-L$ charge of fermions, the 
coupling between electron and f\'eeton can be enhanced.
After showing the parameter space required from the DM stability and cosmic production, we discuss the potential for verifying them in dark matter direct detection experiments. The results show that future experiments, such as 
SuperCDMS, have a sensitivity to reach the f\'eeton DM region  consistent 
with its cosmic production.

\end{abstract}

\section{Introduction}

Many astrophysical observations suggest that $80\%$ of the matter in the universe is dark matter (DM)~\cite{Young:2016ala,Arbey:2021gdg,Cirelli:2024ssz}. However, we still do not know what DM is. A widely accepted conjecture is that DM consists of new particles beyond the framework of the standard model (SM)~\cite{Young:2016ala,Arbey:2021gdg,Cirelli:2024ssz}. 
To be well-motivated, models of DM are typically concise extensions of the SM and capable of addressing certain issues within the SM simultaneously. 
For example, WIMPs predicted from supersymmetry~\cite{Jungman:1995df,Roszkowski:2017nbc}, axions proposed to address the Strong CP problem~\cite{Peccei:1977hh,Weinberg:1977ma,Wilczek:1977pj}. Yet, positive results for the detection of these mainstream candidates are still pending~\cite{ADMX:2020ote,PandaX-4T:2021bab,LZ:2022lsv,XENON:2023cxc}, leading to other possibilities regarding DM.

The $U(1)$ $B-L$ gauge theory is an attractive theory beyond the standard model. This gauge theory 
requires three right-handed neutrinos (RHNs) to cancel the gauge anomalies and its spontaneous breaking induces the super heavy Majorana masses for these RHNs~\cite{FWilczeck:1979}. 
These heavy Majorana RHNs are a key point for generating the observed tiny masses for the active neutrinos (via the seesaw mechanism)~\cite{Minkowski:1977sc, Yanagida:1979as, Yanagida:1979gs, GellMann:1980vs} 
and for the creation of the baryon-number asymmetry in our Universe (by the leptogenesis)~\cite{Fukugita:1986hr, Buchmuller:2005eh}. Furthermore, the $B-L$ gauge boson is an inevitable prediction  in this framework, and
we have stressed recently that it can be a good candidate for DM~\cite{Choi:2020kch} called as the f\'eeton DM~\cite{Lin:2022xbu,Sheng:2023iup}. In this paper, we revisit the physics and parameter space of the f\'eeton dark matter. For the first time, we point out that this  $U(1)$ symmetry can be a linear combination of the $B-L$   
and the SM hypercharge gauge symmetries. With a redefinition of $B-L$ charge of fermions, we find some interesting parameter spaces that not only satisfy the stability and cosmological production requirements of DM but can also be tested in future direct detection experiments such as superCDMS.

\section{Stability of the F\'eeton DM}

The SM can be extended with a $U(1)_{B-L}$ gauge symmetry and its corresponding gauge boson $A'$. The related Lagraingain is, 
\begin{align}
   \mathcal{L} 
  =  (D_\mu \Phi)^\dagger D^{\mu} \Phi - \frac{1}{4} F^{\prime \mu \nu } F^\prime_{\mu \nu} 
    - \lambda \left( |\Phi|^2 - \frac{V_{B-L}^2}{2}\right)^2
     + g_{B-L} q_{B-L} \bar{f} \gamma^\mu f A_\mu^{\prime}.
\end{align}
The covariant derivative $D_\mu$ is defined as $D_\mu \equiv \partial_\mu - i 2 g_{B-L} A_\mu^\prime$ with the gauge coupling $g_{B-L}$, 
as the  Higgs field $\Phi$ has a $B-L$ charge $q_{B-L}=2$.  
The spontaneous symmetry breaking gives the Higgs field $\Phi$ a vacuum expectation value (vev) $V_{B-L}$, its mass 
$m_\phi = \sqrt{2 \lambda} V_{B-L}$, and a gauge boson mass 
$m_{A^\prime} = 2 g_{B-L} V_{B-L}$. 
Any fermions $f$ with a $B-L$ charge $q_{B-L}$ can couple to the gauge boson $A'$ with a universal coupling strength 
$g_{B-L}$. The fermions $f$ include not only the SM quarks with $B-L$ charge $q_{B-L} = 1/3$ and leptons with $q_{B-L} = -1$, but also the heavy RHNs with $q_{B-L} = -1$ to cancel the gauge anomalies. 
Such a framework is self-consistent and well-motivated, 
as the three RHNs are important ingredients for generating the tiny neutrino 
masses via see-saw mechanism~\cite{Minkowski:1977sc, Yanagida:1979as, Yanagida:1979gs,FWilczeck:1979, GellMann:1980vs} and the baryon asymmetry in our Universe via leptongensis~\cite{Fukugita:1986hr, Buchmuller:2005eh}.
In order for the standard leptongensis to work, the vacuum expectation value $V_{B-L}$ should satisfy $V_{B-L} > 3 \times 10^9$~\cite{Fukugita:1986hr,Buchmuller:2005eh,Davidson:2008bu}, which gives the upper bound for the coupling $g_{B-L}$ shown as the dashed purple line in \gfig{fig:FeetonCons}.
This upper limit can be greatly relaxed if we consider some other leptongensis scenarios, such as resonant leptogenesis~\cite{Pilaftsis:2003gt,Pilaftsis:2005rv}.
Furthermore, the gauge boson $A'$ can be a natural candidate of DM in this minimal framework~\cite{Choi:2020kch,Lin:2022xbu,Sheng:2023iup}, called as f\'eeton.

With the coupling to leptons, the f\'eeton with a certain mass will decay. The decay rate is dominant by its decay channel to 
two active neutrinos as, 
\begin{equation}
    \Gamma_{A^\prime}
    \simeq
    \frac{1}{8 \pi} g_{B-L}^2 m_{A^\prime}.
\end{equation}
The required long lifetime for the f\'eeton DM is guaranteed by the extremely small gauge coupling constant $g_{B-L}$. By conservatively assuming that the lifetime of f\'eeton DM excesses ten times of the age of the universe, $\tau_{A^\prime} > 150\,$Gyr, an upper bound on $g_{B-L}$ can be derived, as shown by the black dashed line in \gfig{fig:FeetonCons}. 
Consequently, the grey shaded region is excluded by the requirement of DM stability.
Such a constraint on coupling is notably stringent, with $g_{B-L} \leq 10^{-16}$ even for light f\'eeton with a mass $m_{A'} = 1\,$eV.

Actually, this extra unknown
$U(1)$ gauge symmetry
can be in principle a linear combination of the canonical 
$U(1)_{B-L}$ symmetry and the SM $U(1)_Y$ symmetry, as both them are 
anomaly-free.
All the $B-L$ charges of fermions
can be redefined as $q'_{B-L} \equiv q_{B-L} + \alpha Y$ with a rotation factor $\alpha$~\cite{Lee:2016ief,Hayashi:2024not}.
In such a case, the SM Higgs field is charged under this redefined $U(1)'_{B-L}$ with $q'_{B-L} = \alpha/2$ and gives an additional contribution to f\'eeton mass. 
As a result, we need to re-diagonalize the mass matrix of gauge bosons to obtain their new mass eigenstates and interacting currents.
First, we write the mass terms of the gauge bosons on the basis of the $U(1)_Y$, 
the neutral component of $SU(2)_L$ and the original $U(1)_{B-L}$ as
$V^T \equiv \begin{pmatrix}
    B &
W^3 &
A^\prime
\end{pmatrix}$,
\begin{equation}
\mathcal L_m = 
V^T
\begin{pmatrix}
\frac{1}{8}g'^2 v^2 & -\frac{1}{8} g g' v^2 & 
\frac{1}{8} g' g_{B-L} v^2 \alpha \\
-\frac{1}{8} g g' v^2 & \frac{1}{8}g^2 v^2 & 
-\frac{1}{8} g g_{B-L} v^2 \alpha  \\
\frac{1}{8} g' g_{B-L} v^2 \alpha & 
-\frac{1}{8} g g_{B-L} v^2 \alpha
& 2 g_{B-L}^2 v^2_{B-L} 
+ \frac{1}{8} g^2_{B-L} v^2 \alpha^2
\end{pmatrix}
V
\end{equation}
with $g\,$($g'$) being the gauge couplings of the SM $SU(2)\,$($U(1)_Y$).
Then, we utilize the Weinberg angle, 
$S_w = g^\prime/\sqrt{g^2 + g^{\prime 2}}$ and $C_w = g/\sqrt{g^2 + g^{\prime 2}}$, to 
first diagonalize the $B$ and $W^3$ field into the photon field $\tilde A$ and weak gauge boson $\tilde Z$ as an intermediate step,
\begin{eqnarray}
\begin{pmatrix}
\tilde A \\
\tilde Z \\
\tilde A^\prime
\end{pmatrix}
    &=&
    \left(
    \begin{array}{ccc}
    C_w & S_w & 0\\
    -S_w & C_w & 
    0  \\
    0 & 0 & 1
    \end{array}
    \right)
\left(
\begin{array}{c}
B \\
W^3 \\
A^\prime
\end{array}\right).
\end{eqnarray}
The mass terms in this intermediate basis 
$\tilde V^T \equiv \begin{pmatrix}
\tilde A &
\tilde Z &
\tilde A^\prime
\end{pmatrix}$
becomes, 
\begin{eqnarray}
\mathcal L_m
    &=&
\tilde V^T
\begin{pmatrix}
    0 & 0 & 0\\
    0 & \frac{1}{8} (g^2 + g'^2) v^2 & 
    -\frac{1}{8} \alpha g_{B-L} \sqrt{g^2 + g'^2} v^2 \\
    0 & -\frac{1}{8} \alpha g_{B-L} \sqrt{g^2 + g'^2} v^2 & 2 g_{B-L}^2 v_{B-L}^2 + \frac{1}{8}  g_{B-L}^2  v^2 \alpha^2 
\end{pmatrix}
\tilde V.
\end{eqnarray}
One can see that the SM photon remains massless 
as expected.
After the diagonalization 
of the remaining mass matrix for the 
intermediate $\tilde Z$ and $\tilde A'$, the mass eigenstates $Z$ and $A'$ are defined as, 
\begin{equation}
\begin{pmatrix}
    Z \\
    A^{\prime}
\end{pmatrix}
\equiv
\begin{pmatrix}
    \cos \xi & \sin \xi  \\
    -\sin \xi  & \cos \xi
\end{pmatrix}
\begin{pmatrix}
    \tilde Z \\
    \tilde A^\prime
\end{pmatrix}
\end{equation}
with 
\begin{equation}
\tan 2 \xi
    =
    \frac{-\frac{1}{4} \alpha g_{B-L} \sqrt{g^2 + g'^2} v^2}{\frac{1}{8} (g^2 + g'^2) v^2 - 2 g_{B-L}^2 v_{B-L}^2 - \frac{1}{8}  g_{B-L}^2  v^2 \alpha^2} 
    \sim
    \frac{-2 \alpha g_{B-L}}{\sqrt{g^2+g^{\prime 2}}}.
\end{equation}
With the above rotations, the interaction currents change as well.
The interacting terms of fermions with gauge bosons are,
\begin{equation}
    \mathcal{L}_I
    =
    J^\mu_{em} A_\mu
    +
    \left( \cos \xi J^\mu_{\tilde Z} + \sin \xi J^\mu_{\tilde A^\prime} \right) Z_\mu
    + 
    \left( -\sin \xi J^\mu_{\tilde Z} + \cos \xi J^\mu_{\tilde A^\prime} \right) A^{\prime }_\mu,
\end{equation}
where $J^\mu_{\tilde Z}$ is the $Z$ boson current in SM and 
$J^\mu_{\tilde A'}$ is the original 
f\'eeton current before the redefinition.
Firstly, one can observe that the neutrino-f\'eeton coupling is independent of $\alpha$, 
\begin{align}
\mathcal{L}_\nu
&=
    \left[
    - \frac{g }{2 C_w} \sin \xi
    +g_{B-L} (-1 - \frac{1}{2} \alpha) \cos \xi 
    \right] \bar{\nu}_L \gamma^\mu \nu_L A_\mu^{\prime}
    \approx
    - g_{B-L} \bar{\nu}_L \gamma^\mu \nu_L A_\mu^{\prime},
\end{align}
which means the decay width of f\'eeton and 
the constraint from DM stability remain unchanged
even after the charge redefinition.
However, it is important to note that
the coupling strength between electron and f\'eeton is 
dependent of $\alpha$ as, 
\begin{equation}
\begin{split}
    \mathcal{L}_e 
&=
    \left( g_{B-L} \cos \xi  (-1- \frac{1}{2} \alpha) - g \sin \xi \frac{-\frac{1}{2} + S_w^2}{C_w} 
    \right) \bar{e}_L \gamma^\mu e_L X^m_\mu\\
& \quad +
    \left( g_{B-L} \cos \xi  (-1-  \alpha) - g \sin \xi \frac{ S^2_w}{C_w}
    \right) \bar{e}_R \gamma^\mu e_R X^m_\mu\\
& =
    \left(-1 - C_w^2 \alpha \right) g_{B-L} \bar{e} \gamma^\mu e A'_\mu.
\end{split}
\end{equation}
With a typical $\alpha$, such as $\alpha = 2.6$, the
$B-L$ charge of electron changes from $-1$ to $-3$, leading to a enhancement of the f\'eeton-electron interaction cross section by one order of magnitude.
Due to this redefinition of the $B-L$ charge, the detection capability of this model will also be enhanced by the DM direct detection experiments, as we will discuss later.

\section{Cosmic Production of the F\'eeton DM} 

However, a crucial issue of the f\'eeton DM is its cosmological production, as it is very difficult to produce the f\'eeton in the early universe due to the small gauge coupling constant. 
Nevertheless, two natural mechanisms have been proposed for sufficient vector boson DM productions: 
The decay of cosmic strings associated with  $U(1)$ gauge symmetry breaking~\cite{Long:2019lwl,Kitajima:2022lre} , and the
quantum fluctuations during the inflation~\cite{Graham:2015rva}. We shall discuss their applications to the f\'eeton DM production as follows.

\begin{figure}
    \centering
    \includegraphics[width=0.468
    \textwidth]{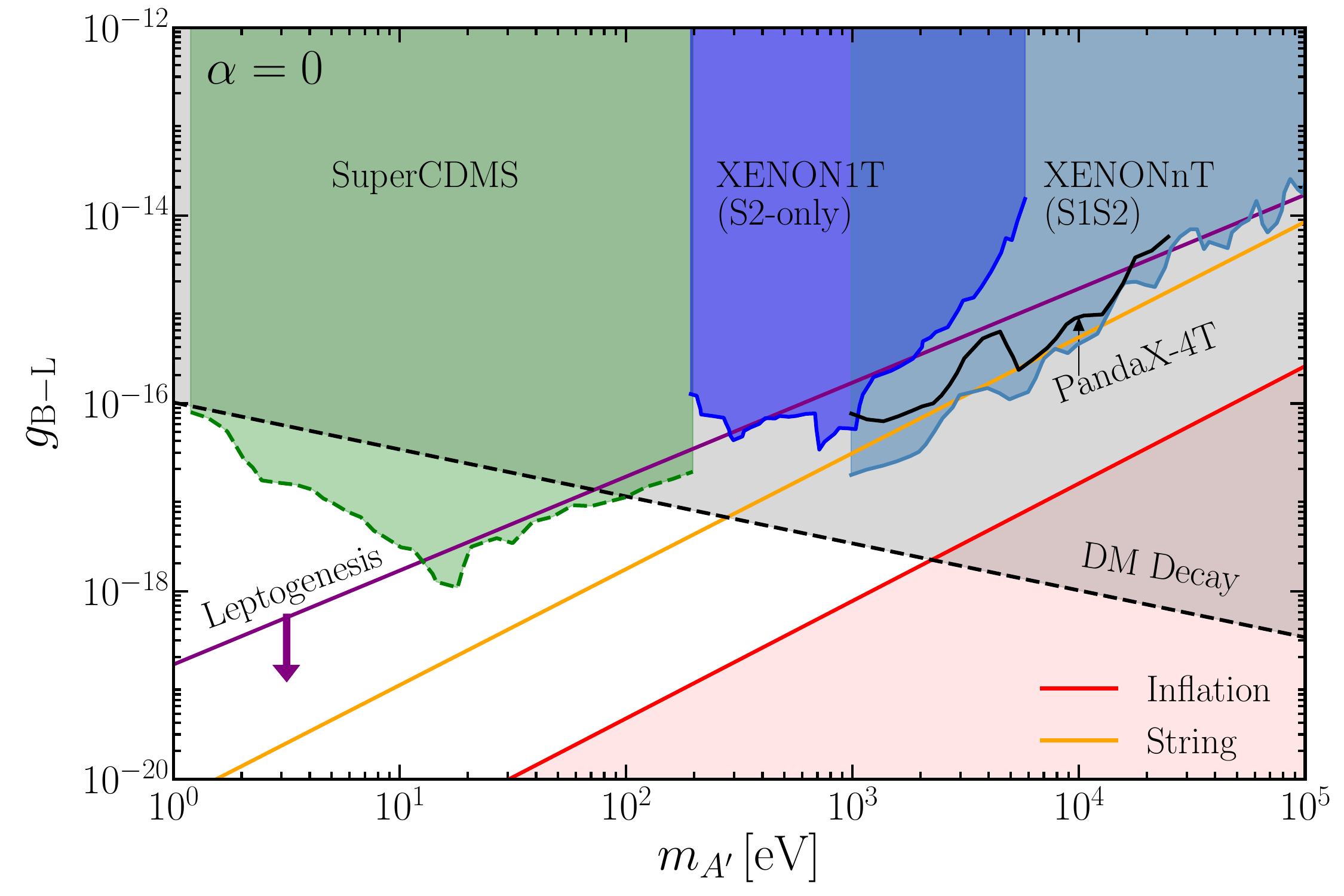}
    \includegraphics[width=0.468
    \textwidth]{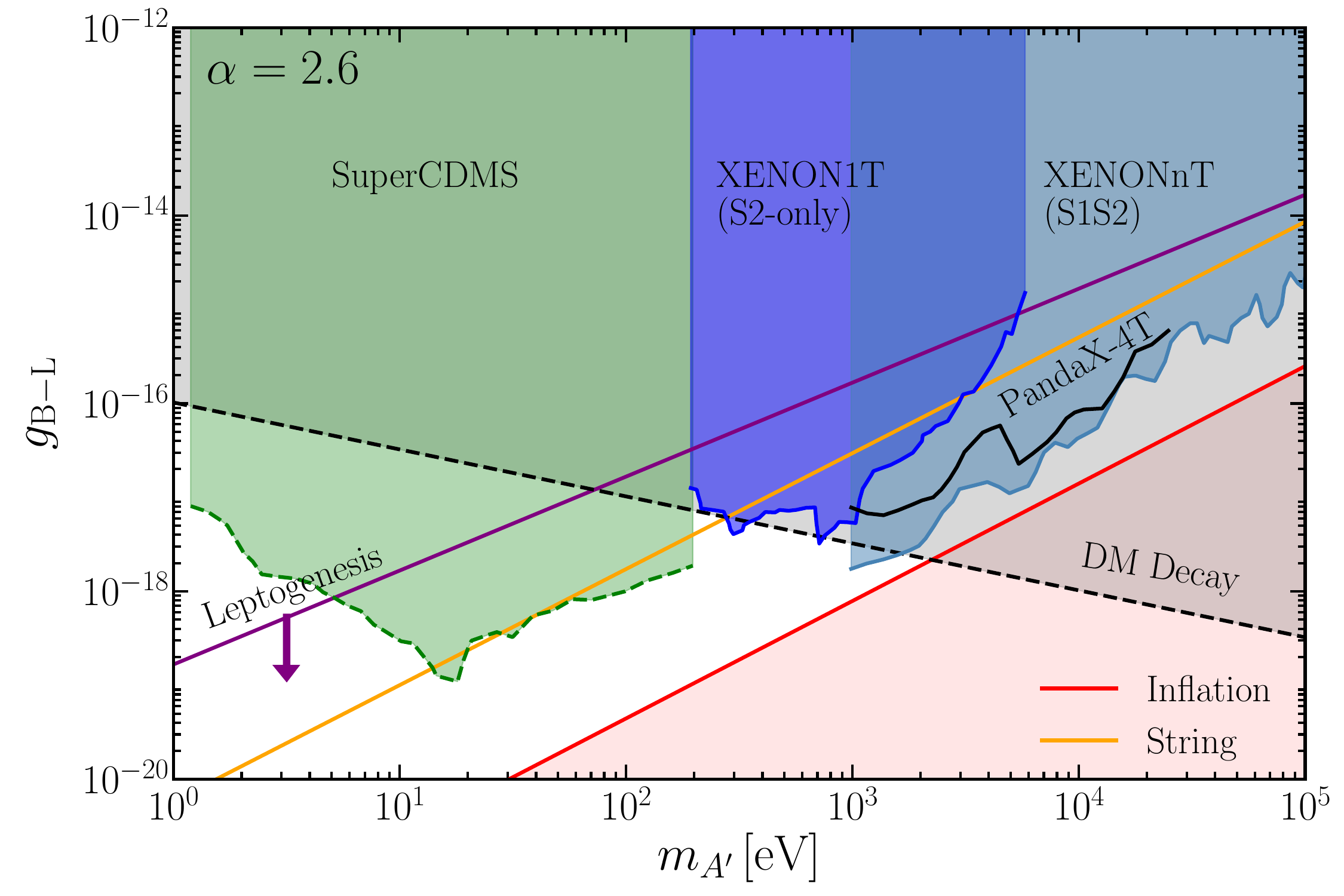}
    \caption{ 
    The parameter space of f\'eeton DM for generating the correct relic density via inflation production  (red shaded region) and cosmic string decay (orange solid line with one order of uncertainty).
    The gray shaded region is excluded since the corresponding lifetime of the f\'eeton DM is shorter than ten times of the age of Universe, $\tau_{\mathfrak{f}} < 150\,$Gyr, due to DM stability. 
    The blue shaded regions are the current constraint from the XENON1T (S2-only) and XENONnT (S1S2) experiments\cite{XENON:2019gfn,XENON:2020rca,XENON:2022ltv} while the green region is the projected limit in future SuperCDMS experiments~\cite{SuperCDMS:2022kse}. The black solid line is the constraint from PandaX-4T~\cite{PandaX:2024cic}.
    All these direct detection constraints
    are dependent of the charge redefinition 
    parameter $\alpha$.
    The left panel corresponds to the case of $\alpha = 0$ and the right panel the case of $\alpha = 2.6$.
     }
    \label{fig:FeetonCons}
\end{figure}

The f\'eeton DM, as the gauge boson of the $U(1)_{B-L}$ gauge symmetry, gains its mass from the Higgs mechanism.
The dark Higgs field $\Phi$
obtains a vacuum expectation value $V_{B-L}$, and the $U(1)_{B-L}$ symmetry is spontaneously broken.  
Once the symmetry breaking happens after inflation, cosmic string networks appear~\cite{Kitajima:2023vre} and their subsequent decay produce 
the f\'eeton DM relic abundance $ \Omega_f $,
\begin{equation}
    \Omega_f 
    \simeq 
    0.12
    \left(\frac{\xi}{5.7}\right)
    \left(\frac{m_{A^\prime}}{1 \,\mathrm{keV}}\right)^{1 / 2}
    \left(\frac{V_{B-L}}{1.6 \times 10^{10}\, \mathrm{GeV}}\right)^2,
\end{equation}
can be consistent with the observed value.
The average string number per Hubble volume 
$\xi$ can be obtained from simulations~\cite{Gorghetto:2018myk,Gorghetto:2020qws},
\begin{equation}
    \xi  = 0.15 \log \left(\frac{m_\Phi}{m_{A^\prime}}\right)
    =
    0.15 \log \left( \frac{\sqrt{2 \lambda} V_{B-L}}{m_{A^\prime}}\right)
\label{cosmisString}
\end{equation}
The corresponding parameters that generate the correct relic density by taking $\lambda = 1/2$ are shown as the yellow solid line in \gfig{fig:FeetonCons}. Note that there is an $\mathcal{O}(1)$ uncertainty in the relic density calculation, which arises from the uncertainties in the cosmic string simulation \footnote{We would acknowledge the communication with N. Kitajima and K. Nakayama on this point.}. This means that the parameter space is actually a band centered around the yellow line with a width of one order.

The other possibility for the f\'eeton DM production is the production of massive vector boson from the quantum fluctuations during inflation.
The massive vector field is initially sub-horizon
and evolves to become super-horizon modes during inflation. 
Subsequently, they re-enter the horizon during radiation domination to be cosmic relics.
In such a scenario, the relic density of f\'eeton is precisely determined by its mass and 
the Hubble scale of inflation $H_I$~\cite{Graham:2015rva}, 
\begin{equation}
    \Omega_{A^\prime}
    =
    0.3
    \sqrt{\frac{m_{A^\prime}}{1 \,
    {\rm keV}} }
    \left(\frac{H_I}{10^{12}\, {\rm GeV}}\right)^2.
\end{equation}
It is required that the spontaneous breaking of $B-L$ gauge symmetry occurs before or during inflation, leading to $V_{B-L} > H_I$. Taking $H_I$ as a lower bound for $V_{B-L}$, 
the above relic density formula has an same scaling of $m_{A^\prime}$ and $V_{B-L}$ as the cosmic string case \geqn{cosmisString}.
As a result, the upper bound line for f\'eeton coupling $g_{B-L}$ as a function of $m_{A'}$ to generate the right relic density (red solid) is 
parallel to that of the cosmic string production but with around $2$ orders smaller. The larger vev $V_{B-L}$, or equivalently, smaller coupling $g_{B-L}$ area (red shaded region) can also generate the correct DM relic density.

\section{Direct Detection of the F\'eeton DM}

Although the requirements for stability and relic density have predicted a very small f\'eeton DM coupling, our scenario can still be tested by future direct detection experiments.

The f\'eeton DM can be absorbed 
by the electron targets in the DM direct detection experiments via the leading process $ e + A' \rightarrow e + \gamma$~\cite{Pospelov:2008jk,An:2013yua,An:2014twa,Hochberg:2016ajh,Bloch:2016sjj,Hochberg:2016sqx,An:2020bxd,Caputo:2021eaa,Ge:2022ius}. With the total 
DM mass transferred into the kinetic energy of the final states, such an absorption 
process allows a lower mass threshold for DM detection~\cite{Dror:2020czw,Ge:2022ius,PandaX:2022ood}.
In the Xenon-based DM direct detection experiments, such as Xenon~\cite{XENON:2019gfn,XENON:2020rca,XENON:2022ltv}, PandaX~\cite{PandaX:2024cic,PandaX:2024kjp}, and LZ~\cite{LZ:2021xov,LZ:2023poo}, the most outer layer electrons of 
Xenon atom has a binding energy of $\sim 10\,$eV. 
Due to the energy conservation, the f\'eeton DM with a mass larger than $10\,$eV
is allowed to be absorbed and ionized the electrons.
Therefore, these experiments place constraints on f\'eeton coupling above the f\'eeton mass $m_{A'} \sim 10\,$eV, and become the strongest terrestrial experimental constraints beyond $\sim 200\,$eV. The stringent constraint from XENON1T (S2 only)~\cite{XENON:2019gfn} and XENONnT (S1S2) experiments~\cite{XENON:2022ltv} are shown as the blue regions in the left panel of \gfig{fig:FeetonCons}. The PandaX-4T experiment currently has a better S2-only sensitivity~\cite{PandaX:2022xqx,PandaX:2024muv}. 
Although the experimental testability has reached the Leptogenesis and string production lines,
such a parameter space is already disfavored from the stability requirement for DM.

The next-generation SuperCDMS experiment, which will be developed
in SNOLAB~\cite{SuperCDMS:2022kse}, is expected to have a 
lower detection threshold and 
better projected sensitivity in the low mass range.
The detection threshold of a phonon detector is determined by the band gap of the materials it is composed of. For gemanium (Ge), which is planned to be used in the SuperCDMS project, the threshold can be lowered to 0.67 eV.
The projected sensitivity of the SuperCDMS experiment for the f\'eeton absorption process is shown by the green region in the \gfig{fig:FeetonCons}. 
It provides the strongest constraints on f\'eeton coupling in the mass range from $\sim 1\,$eV to 
$\sim 200\,$eV. 
This projected constraint can reach the f\'eeton DM region  
below the black dashed line, and surpass the upper limit of canonical leptogenesis as well in the f\'eeton mass range $\sim (10, 100)\,$eV.

All these direct detection constraints become stronger if the 
$U(1)_{B-L}$ symmetry contains a 
linear combination with the SM $U(1)_Y$ symmetry, where $\alpha \neq 0$.
As previously stated, if $\alpha = 2.6$, the charge of the electron, or equivalently, its coupling with f\'eeton, increases by three times. 
As a result, the cross section of f\'eeton absorption increases by a factor of $\sim 10$, thereby strengthening the constraints on the f\'eeton model with this $\alpha$ by an order of magnitude in direct detection experiments. We show these
enhanced limits in the right panel of \gfig{fig:FeetonCons}.
One can see the future SuperCDMS can test the f\'eeton DM parameter space 
which is consistent with the cosmic string production in the mass range 
$m_{A'} \subset (10,200)\,$eV. The future Xenon-based detectors, with tenfold increased exposure and reduced backgrounds, can even get a better sensitivity~\cite{Wang:2023wrr,XLZD:2024gxx}, having the possibility to
test the inflation production of f\'eeton DM with a mass around $m_{A'} \sim 10^3\,$eV.

\section{Conclusions and Discussions}

\textit{What is dark matter?} This is one of the most important questions in modern physics. There is no universal method for the detection of DM; 
it must rely on the mass and interaction type of DM predicted by DM models. Therefore, clearly defined, theoretically well-motivated DM candidates are of great importance. Currently, our strategy is to search for DM candidates that can address issues in the SM. Among the most popular are WIMPs and the axion.
WIMPs can be predicted by a broad range of new physics models. One well-motivated class is supersymmetry model, which was initially proposed to unify all interactions and can address the hierarchy problem in the SM~\cite{Dimopoulos:1981au,Dimopoulos:1981zb,Martin:1997ns}. However, so far, no supersymmetric particles have been discovered at colliders, and direct detection of DM has placed strong constraints on WIMPs. 
On the other hand, the proposal of the axion aimed to theoretically solve the strong CP problem, but it also suffers from more severe high-quality problem~\cite{Kamionkowski:1992mf,Holman:1992us}. 
And there are alternative solutions to the strong CP problem that do not require the introduction of the axion~\cite{Nelson:1983zb,Barr:1984qx,Nelson:1984hg,Choi:2019omm,Feruglio:2023uof,Feruglio:2024ytl,Liang:2024wbb}.

Compared to theoretical puzzles like the hierarchy problem and the strong CP problem, phenomena beyond the SM 
such as neutrino masses and baryon asymmetry pose more concrete phenomenological questions. 
Within the framework of the seesaw mechanism, which introduces three heavy Majorana RHNs, both of these issues can be effectively addressed. 
Moreover, this framework can also motivate the existence of a new $B-L$ $U(1)$ gauge symmetry,  incorporating the corresponding gauge boson $A'$.
As long as the gauge coupling is sufficiently small to ensure the stability of $A'$, it can naturally become DM. We call it f\'eeton DM. This DM candidate predicted from the simple and elegant extension of SM deserves more attention.

In this paper, we revisit the physics of f\'eeton DM, and 
show the viable parameter space required from the stability of DM.
After that, we review the cosmic production of f\'eeton DM via cosmic string decay 
and inflation.
In this canonical f\'eeton DM model, the most parameter space the
DM direct detection can constrain is already excluded due to the DM decay. The future SuperCDMS, however, can reach the f\'eeton DM region, which is consistent with the 
canonical leptogenesis with $V_{B-L} > 3 \times 10^9\,$GeV. More importantly, we first point out the freedom of the definition 
of $B-L$ charge, $q'_{B-L} \rightarrow q_{B-L} + \alpha Y$, since this $U(1)$ gauge symmetry can be a linear combination of the $B-L$  and the SM hypercharge gauge symmetries.
With this redefinition, the coupling between electron and f\'eeton can be 
enhanced, leading to a stronger limit from DM direct detection. In the 
case of $\alpha = 2.6$, as the example in our paper, all the direct detection 
constrained is enhanced by one order of magnitude. In such a scenario, the future 
SuperCDMS will have the capability to test the cosmic string production of 
f\'eeton DM in the mass range $m_{A'} \subset (10,200)\,$eV while the future Xenon-based 
detectors, such as the Xenon-nT, PandaX, and LZ, have a chance to test the inflation production of f\'eeton DM with a mass 
$m_{A'} \sim 1\,$keV.

For the f\'eeton DM lighter than $1\,$eV, there are also some experimental 
proposals, such as the Josephson junction current~\cite{Cheng:2024yrn,Cheng:2024zde} and molecule excitations~\cite{Arvanitaki:2017nhi}, to detect them. However, the predicted coupling from DM production
is too small to be reached. In even lighter mass ranges, like $m_{A'} < 10^{-5}\,$eV, fifth force experiments have imposed very strong constraints~\cite{Adelberger:2009zz,Wagner:2012ui,LIGOScientific:2021ffg,Miller:2023kkd,Frerick:2023xnf}.

Finally, we should comment on the serious difference between the f\'eeton DM model and the dark photon DM model. The dark photon DM couples to the electron only through the kinetic mixing term between the dark photon and the SM photon~\cite{Fabbrichesi:2020wbt}. Thus, we do not have any prediction for the detection experiments as its kinetic mixing $\epsilon$ is a free parameter, which is independent from the DM mass and the vaccum expectation value of the dark gauge symmetry breaking scale. It is crucial the f\'eeton mass is given by $m_{A'}=2 g_{B-L} V_{B-L}$. This is the reason why we have a certain prediction for the detection of the f\'eeton DM in the future direct detection experiments.

\section*{Acknowledgements}

We thank N. Kitajima and K. Nakayama for discussion on the f\'eeton production from cosmic string decays. We also thank Jianglai Liu for useful discussion on the f\'eeton direct detection.
J. S. is supported by the National Natural Science Foundation of China (Nos. 12375101, 12425506, 12090060, 12090064) and the SJTU Double First Class start-up fund WF220442604. 
T. T. Y. was supported also by the Natural Science Foundation of China (NSFC)
under Grant No. 12175134, MEXT KAKENHI Grants No. 24H02244, and World Premier International Research Center Initiative
(WPI Initiative), MEXT, Japan.

\providecommand{\href}[2]{#2}\begingroup\raggedright\endgroup


\begin{thebibliography}{10}
	
	\bibitem{Young:2016ala}
	B.-L. Young, ``{\it {A survey of dark matter and related topics in
			cosmology}},'' \href{http://dx.doi.org/10.1007/s11467-016-0583-4}{{\em Front.
			Phys. (Beijing)} {\bfseries 12} no.~2, (2017) 121201}. [Erratum:
	Front.Phys.(Beijing) 12, 121202 (2017)].
	
	\bibitem{Arbey:2021gdg}
	A.~Arbey and F.~Mahmoudi, ``{\it {Dark matter and the early Universe: a
			review}},'' \href{http://dx.doi.org/10.1016/j.ppnp.2021.103865}{{\em Prog.
			Part. Nucl. Phys.} {\bfseries 119} (2021) 103865},
	[\href{http://arxiv.org/abs/2104.11488}{{\ttfamily arXiv:2104.11488}}
	[hep-ph]].
	
	\bibitem{Cirelli:2024ssz}
	M.~Cirelli, A.~Strumia, and J.~Zupan, ``{\it {Dark Matter}},''
	[\href{http://arxiv.org/abs/2406.01705}{{\ttfamily arXiv:2406.01705}}
	[hep-ph]].
	
	\bibitem{Jungman:1995df}
	G.~Jungman, M.~Kamionkowski, and K.~Griest, ``{\it {Supersymmetric dark
			matter}},'' \href{http://dx.doi.org/10.1016/0370-1573(95)00058-5}{{\em Phys.
			Rept.} {\bfseries 267} (1996) 195--373},
	[\href{http://arxiv.org/abs/hep-ph/9506380}{{\ttfamily
			arXiv:hep-ph/9506380}}].
	
	\bibitem{Roszkowski:2017nbc}
	L.~Roszkowski, E.~M. Sessolo, and S.~Trojanowski, ``{\it {WIMP dark matter
			candidates and searches\textemdash{}current status and future prospects}},''
	\href{http://dx.doi.org/10.1088/1361-6633/aab913}{{\em Rept. Prog. Phys.}
		{\bfseries 81} no.~6, (2018) 066201},
	[\href{http://arxiv.org/abs/1707.06277}{{\ttfamily arXiv:1707.06277}}
	[hep-ph]].
	
	\bibitem{Peccei:1977hh}
	R.~D. Peccei and H.~R. Quinn, ``{\it {CP Conservation in the Presence of
			Instantons}},'' \href{http://dx.doi.org/10.1103/PhysRevLett.38.1440}{{\em
			Phys. Rev. Lett.} {\bfseries 38} (1977) 1440--1443}.
	
	\bibitem{Weinberg:1977ma}
	S.~Weinberg, ``{\it {A New Light Boson?}},''
	\href{http://dx.doi.org/10.1103/PhysRevLett.40.223}{{\em Phys. Rev. Lett.}
		{\bfseries 40} (1978) 223--226}.
	
	\bibitem{Wilczek:1977pj}
	F.~Wilczek, ``{\it {Problem of Strong $P$ and $T$ Invariance in the Presence of
			Instantons}},'' \href{http://dx.doi.org/10.1103/PhysRevLett.40.279}{{\em
			Phys. Rev. Lett.} {\bfseries 40} (1978) 279--282}.
	
	\bibitem{ADMX:2020ote}
	{\bfseries ADMX} Collaboration, R.~Khatiwada {\em et~al.}, ``{\it {Axion Dark
			Matter Experiment: Detailed design~and operations}},''
	\href{http://dx.doi.org/10.1063/5.0037857}{{\em Rev. Sci. Instrum.}
		{\bfseries 92} no.~12, (2021) 124502},
	[\href{http://arxiv.org/abs/2010.00169}{{\ttfamily arXiv:2010.00169}}
	[astro-ph.IM]].
	
	\bibitem{PandaX-4T:2021bab}
	{\bfseries PandaX-4T} Collaboration, Y.~Meng {\em et~al.}, ``{\it {Dark Matter
			Search Results from the PandaX-4T Commissioning Run}},''
	\href{http://dx.doi.org/10.1103/PhysRevLett.127.261802}{{\em Phys. Rev.
			Lett.} {\bfseries 127} no.~26, (2021) 261802},
	[\href{http://arxiv.org/abs/2107.13438}{{\ttfamily arXiv:2107.13438}}
	[hep-ex]].
	
	\bibitem{LZ:2022lsv}
	{\bfseries LZ} Collaboration, J.~Aalbers {\em et~al.}, ``{\it {First Dark
			Matter Search Results from the LUX-ZEPLIN (LZ) Experiment}},''
	\href{http://dx.doi.org/10.1103/PhysRevLett.131.041002}{{\em Phys. Rev.
			Lett.} {\bfseries 131} no.~4, (2023) 041002},
	[\href{http://arxiv.org/abs/2207.03764}{{\ttfamily arXiv:2207.03764}}
	[hep-ex]].
	
	\bibitem{XENON:2023cxc}
	{\bfseries XENON} Collaboration, E.~Aprile {\em et~al.}, ``{\it {First Dark
			Matter Search with Nuclear Recoils from the XENONnT Experiment}},''
	\href{http://dx.doi.org/10.1103/PhysRevLett.131.041003}{{\em Phys. Rev.
			Lett.} {\bfseries 131} no.~4, (2023) 041003},
	[\href{http://arxiv.org/abs/2303.14729}{{\ttfamily arXiv:2303.14729}}
	[hep-ex]].
	
	\bibitem{FWilczeck:1979}
	F.~Wilczeck, ``{\it {Proceedings: Lepton-Photon Conference (Fermilab, Aug
			1979)}},'' {\em Conf. Proc. C790885} (1979) .
	
	\bibitem{Minkowski:1977sc}
	P.~Minkowski, ``{\it {$\mu \to e\gamma$ at a Rate of One Out of $10^{9}$ Muon
			Decays?}},''
	\href{http://dx.doi.org/10.1016/0370-2693(77)90435-X}{{\em Phys. Lett.}
		{\bfseries 67B} (1977) 421--428}.
	
	\bibitem{Yanagida:1979as}
	T.~Yanagida, ``{\it {Horizontal gauge symmetry and masses of neutrinos}},''
	{\em {Proceedings: Workshop on the Unified Theories and the Baryon Number in
			the Universe}, Conf. Proc.} {\bfseries C7902131} (1979) 95--99.
	
	\bibitem{Yanagida:1979gs}
	T.~Yanagida, ``{\it {Horizontal Symmetry and Mass of the Top Quark}},''
	\href{http://dx.doi.org/10.1103/PhysRevD.20.2986}{{\em Phys. Rev. D}
		{\bfseries 20} (1979) 2986}.
	
	\bibitem{GellMann:1980vs}
	M.~Gell-Mann, P.~Ramond, and R.~Slansky, ``{\it {Complex Spinors and Unified
			Theories}},'' {\em Conf. Proc.} {\bfseries C790927} (1979) 315--321,
	[\href{http://arxiv.org/abs/1306.4669}{{\ttfamily arXiv:1306.4669}} [hep-th]].
	
	\bibitem{Fukugita:1986hr}
	M.~Fukugita and T.~Yanagida, ``{\it {Baryogenesis Without Grand
			Unification}},'' \href{http://dx.doi.org/10.1016/0370-2693(86)91126-3}{{\em
			Phys. Lett. B} {\bfseries 174} (1986) 45--47}.
	
	\bibitem{Buchmuller:2005eh}
	W.~Buchmuller, R.~D. Peccei, and T.~Yanagida, ``{\it {Leptogenesis as the
			origin of matter}},''
	\href{http://dx.doi.org/10.1146/annurev.nucl.55.090704.151558}{{\em Ann. Rev.
			Nucl. Part. Sci.} {\bfseries 55} (2005) 311--355},
	[\href{http://arxiv.org/abs/hep-ph/0502169}{{\ttfamily
			arXiv:hep-ph/0502169}}].
	
	\bibitem{Choi:2020kch}
	G.~Choi, T.~T. Yanagida, and N.~Yokozaki, ``{\it {Feebly interacting
			$U(1)_{B-L}$ gauge boson warm dark matter and XENON1T anomaly}},''
	\href{http://dx.doi.org/10.1016/j.physletb.2020.135836}{{\em Phys. Lett. B}
		{\bfseries 810} (2020) 135836},
	[\href{http://arxiv.org/abs/2007.04278}{{\ttfamily arXiv:2007.04278}}
	[hep-ph]].
	
	\bibitem{Lin:2022xbu}
	W.~Lin, L.~Visinelli, D.~Xu, and T.~T. Yanagida, ``{\it {Neutrino astronomy as
			a probe of physics beyond the Standard Model: Decay of sub-MeV B-L gauge
			boson dark matter}},''
	\href{http://dx.doi.org/10.1103/PhysRevD.106.075011}{{\em Phys. Rev. D}
		{\bfseries 106} no.~7, (2022) 075011},
	[\href{http://arxiv.org/abs/2202.04496}{{\ttfamily arXiv:2202.04496}}
	[hep-ph]].
	
	\bibitem{Sheng:2023iup}
	J.~Sheng, Y.~Cheng, W.~Lin, and T.~T. Yanagida, ``{\it {F\'eeton (B-L gauge
			boson) dark matter for the 511-keV gamma-ray excess and the prediction of
			low-energy neutrino flux*}},''
	\href{http://dx.doi.org/10.1088/1674-1137/ad4af3}{{\em Chin. Phys. C}
		{\bfseries 48} no.~8, (2024) 083104},
	[\href{http://arxiv.org/abs/2310.05420}{{\ttfamily arXiv:2310.05420}}
	[hep-ph]].
	
	\bibitem{Davidson:2008bu}
	S.~Davidson, E.~Nardi, and Y.~Nir, ``{\it {Leptogenesis}},''
	\href{http://dx.doi.org/10.1016/j.physrep.2008.06.002}{{\em Phys. Rept.}
		{\bfseries 466} (2008) 105--177},
	[\href{http://arxiv.org/abs/0802.2962}{{\ttfamily arXiv:0802.2962}}
	[hep-ph]].
	
	\bibitem{Pilaftsis:2003gt}
	A.~Pilaftsis and T.~E.~J. Underwood, ``{\it {Resonant leptogenesis}},''
	\href{http://dx.doi.org/10.1016/j.nuclphysb.2004.05.029}{{\em Nucl. Phys. B}
		{\bfseries 692} (2004) 303--345},
	[\href{http://arxiv.org/abs/hep-ph/0309342}{{\ttfamily
			arXiv:hep-ph/0309342}}].
	
	\bibitem{Pilaftsis:2005rv}
	A.~Pilaftsis and T.~E.~J. Underwood, ``{\it {Electroweak-scale resonant
			leptogenesis}},'' \href{http://dx.doi.org/10.1103/PhysRevD.72.113001}{{\em
			Phys. Rev. D} {\bfseries 72} (2005) 113001},
	[\href{http://arxiv.org/abs/hep-ph/0506107}{{\ttfamily
			arXiv:hep-ph/0506107}}].
	
	\bibitem{Lee:2016ief}
	H.-S. Lee and S.~Yun, ``{\it {Mini force: The $(B-L)+xY$ gauge interaction with
			a light mediator}},''
	\href{http://dx.doi.org/10.1103/PhysRevD.93.115028}{{\em Phys. Rev. D}
		{\bfseries 93} no.~11, (2016) 115028},
	[\href{http://arxiv.org/abs/1604.01213}{{\ttfamily arXiv:1604.01213}}
	[hep-ph]].
	
	\bibitem{Hayashi:2024not}
	T.~Hayashi, S.~Matsumoto, Y.~Watanabe, and T.~T. Yanagida, ``{\it {F\'eeton
			dark matter above the $e^-e^+$ threshold}},''
	[\href{http://arxiv.org/abs/2408.12155}{{\ttfamily arXiv:2408.12155}}
	[hep-ph]].
	
	\bibitem{Long:2019lwl}
	A.~J. Long and L.-T. Wang, ``{\it {Dark Photon Dark Matter from a Network of
			Cosmic Strings}},'' \href{http://dx.doi.org/10.1103/PhysRevD.99.063529}{{\em
			Phys. Rev. D} {\bfseries 99} no.~6, (2019) 063529},
	[\href{http://arxiv.org/abs/1901.03312}{{\ttfamily arXiv:1901.03312}}
	[hep-ph]].
	
	\bibitem{Kitajima:2022lre}
	N.~Kitajima and K.~Nakayama, ``{\it {Dark photon dark matter from cosmic
			strings and gravitational wave background}},''
	\href{http://dx.doi.org/10.1007/JHEP08(2023)068}{{\em JHEP} {\bfseries 08}
		(2023) 068}, [\href{http://arxiv.org/abs/2212.13573}{{\ttfamily
			arXiv:2212.13573}} [hep-ph]].
	
	\bibitem{Graham:2015rva}
	P.~W. Graham, J.~Mardon, and S.~Rajendran, ``{\it {Vector Dark Matter from
			Inflationary Fluctuations}},''
	\href{http://dx.doi.org/10.1103/PhysRevD.93.103520}{{\em Phys. Rev. D}
		{\bfseries 93} no.~10, (2016) 103520},
	[\href{http://arxiv.org/abs/1504.02102}{{\ttfamily arXiv:1504.02102}}
	[hep-ph]].
	
	\bibitem{Kitajima:2023vre}
	N.~Kitajima and K.~Nakayama, ``{\it {Nanohertz gravitational waves from cosmic
			strings and dark photon dark matter}},''
	\href{http://dx.doi.org/10.1016/j.physletb.2023.138213}{{\em Phys. Lett. B}
		{\bfseries 846} (2023) 138213},
	[\href{http://arxiv.org/abs/2306.17390}{{\ttfamily arXiv:2306.17390}}
	[hep-ph]].
	
	\bibitem{Gorghetto:2018myk}
	M.~Gorghetto, E.~Hardy, and G.~Villadoro, ``{\it {Axions from Strings: the
			Attractive Solution}},''
	\href{http://dx.doi.org/10.1007/JHEP07(2018)151}{{\em JHEP} {\bfseries 07}
		(2018) 151}, [\href{http://arxiv.org/abs/1806.04677}{{\ttfamily
			arXiv:1806.04677}} [hep-ph]].
	
	\bibitem{Gorghetto:2020qws}
	M.~Gorghetto, E.~Hardy, and G.~Villadoro, ``{\it {More axions from strings}},''
	\href{http://dx.doi.org/10.21468/SciPostPhys.10.2.050}{{\em SciPost Phys.}
		{\bfseries 10} no.~2, (2021) 050},
	[\href{http://arxiv.org/abs/2007.04990}{{\ttfamily arXiv:2007.04990}}
	[hep-ph]].
	
	\bibitem{XENON:2019gfn}
	{\bfseries XENON} Collaboration, E.~Aprile {\em et~al.}, ``{\it {Light Dark
			Matter Search with Ionization Signals in XENON1T}},''
	\href{http://dx.doi.org/10.1103/PhysRevLett.123.251801}{{\em Phys. Rev.
			Lett.} {\bfseries 123} no.~25, (2019) 251801},
	[\href{http://arxiv.org/abs/1907.11485}{{\ttfamily arXiv:1907.11485}}
	[hep-ex]].
	
	\bibitem{XENON:2020rca}
	{\bfseries XENON} Collaboration, E.~Aprile {\em et~al.}, ``{\it {Excess
			electronic recoil events in XENON1T}},''
	\href{http://dx.doi.org/10.1103/PhysRevD.102.072004}{{\em Phys. Rev. D}
		{\bfseries 102} no.~7, (2020) 072004},
	[\href{http://arxiv.org/abs/2006.09721}{{\ttfamily arXiv:2006.09721}}
	[hep-ex]].
	
	\bibitem{XENON:2022ltv}
	{\bfseries XENON} Collaboration, E.~Aprile {\em et~al.}, ``{\it {Search for New
			Physics in Electronic Recoil Data from XENONnT}},''
	\href{http://dx.doi.org/10.1103/PhysRevLett.129.161805}{{\em Phys. Rev.
			Lett.} {\bfseries 129} no.~16, (2022) 161805},
	[\href{http://arxiv.org/abs/2207.11330}{{\ttfamily arXiv:2207.11330}}
	[hep-ex]].
	
	\bibitem{SuperCDMS:2022kse}
	{\bfseries SuperCDMS} Collaboration, M.~F. Albakry {\em et~al.}, ``{\it {A
			Strategy for Low-Mass Dark Matter Searches with Cryogenic Detectors in the
			SuperCDMS SNOLAB Facility}},'' in {\em {Snowmass 2021}}.
	\newblock 3, 2022.
	\newblock [\href{http://arxiv.org/abs/2203.08463}{{\ttfamily arXiv:2203.08463}}
	[physics.ins-det]].
	
	\bibitem{PandaX:2024cic}
	{\bfseries PandaX} Collaboration, X.~Zeng {\em et~al.}, ``{\it {Exploring New
			Physics with PandaX-4T Low Energy Electronic Recoil Data}},''
	[\href{http://arxiv.org/abs/2408.07641}{{\ttfamily arXiv:2408.07641}}
	[hep-ex]].
	
	\bibitem{Pospelov:2008jk}
	M.~Pospelov, A.~Ritz, and M.~B. Voloshin, ``{\it {Bosonic super-WIMPs as
			keV-scale dark matter}},''
	\href{http://dx.doi.org/10.1103/PhysRevD.78.115012}{{\em Phys. Rev. D}
		{\bfseries 78} (2008) 115012},
	[\href{http://arxiv.org/abs/0807.3279}{{\ttfamily arXiv:0807.3279}}
	[hep-ph]].
	
	\bibitem{An:2013yua}
	H.~An, M.~Pospelov, and J.~Pradler, ``{\it {Dark Matter Detectors as Dark
			Photon Helioscopes}},''
	\href{http://dx.doi.org/10.1103/PhysRevLett.111.041302}{{\em Phys. Rev.
			Lett.} {\bfseries 111} (2013) 041302},
	[\href{http://arxiv.org/abs/1304.3461}{{\ttfamily arXiv:1304.3461}}
	[hep-ph]].
	
	\bibitem{An:2014twa}
	H.~An, M.~Pospelov, J.~Pradler, and A.~Ritz, ``{\it {Direct Detection
			Constraints on Dark Photon Dark Matter}},''
	\href{http://dx.doi.org/10.1016/j.physletb.2015.06.018}{{\em Phys. Lett. B}
		{\bfseries 747} (2015) 331--338},
	[\href{http://arxiv.org/abs/1412.8378}{{\ttfamily arXiv:1412.8378}}
	[hep-ph]].
	
	\bibitem{Hochberg:2016ajh}
	Y.~Hochberg, T.~Lin, and K.~M. Zurek, ``{\it {Detecting Ultralight Bosonic Dark
			Matter via Absorption in Superconductors}},''
	\href{http://dx.doi.org/10.1103/PhysRevD.94.015019}{{\em Phys. Rev. D}
		{\bfseries 94} no.~1, (2016) 015019},
	[\href{http://arxiv.org/abs/1604.06800}{{\ttfamily arXiv:1604.06800}}
	[hep-ph]].
	
	\bibitem{Bloch:2016sjj}
	I.~M. Bloch, R.~Essig, K.~Tobioka, T.~Volansky, and T.-T. Yu, ``{\it {Searching
			for Dark Absorption with Direct Detection Experiments}},''
	\href{http://dx.doi.org/10.1007/JHEP06(2017)087}{{\em JHEP} {\bfseries 06}
		(2017) 087}, [\href{http://arxiv.org/abs/1608.02123}{{\ttfamily
			arXiv:1608.02123}} [hep-ph]].
	
	\bibitem{Hochberg:2016sqx}
	Y.~Hochberg, T.~Lin, and K.~M. Zurek, ``{\it {Absorption of light dark matter
			in semiconductors}},''
	\href{http://dx.doi.org/10.1103/PhysRevD.95.023013}{{\em Phys. Rev. D}
		{\bfseries 95} no.~2, (2017) 023013},
	[\href{http://arxiv.org/abs/1608.01994}{{\ttfamily arXiv:1608.01994}}
	[hep-ph]].
	
	\bibitem{An:2020bxd}
	H.~An, M.~Pospelov, J.~Pradler, and A.~Ritz, ``{\it {New limits on dark photons
			from solar emission and keV scale dark matter}},''
	\href{http://dx.doi.org/10.1103/PhysRevD.102.115022}{{\em Phys. Rev. D}
		{\bfseries 102} (2020) 115022},
	[\href{http://arxiv.org/abs/2006.13929}{{\ttfamily arXiv:2006.13929}}
	[hep-ph]].
	
	\bibitem{Caputo:2021eaa}
	A.~Caputo, A.~J. Millar, C.~A.~J. O'Hare, and E.~Vitagliano, ``{\it {Dark
			photon limits: A handbook}},''
	\href{http://dx.doi.org/10.1103/PhysRevD.104.095029}{{\em Phys. Rev. D}
		{\bfseries 104} no.~9, (2021) 095029},
	[\href{http://arxiv.org/abs/2105.04565}{{\ttfamily arXiv:2105.04565}}
	[hep-ph]].
	
	\bibitem{Ge:2022ius}
	S.-F. Ge, X.-G. He, X.-D. Ma, and J.~Sheng, ``{\it {Revisiting the fermionic
			dark matter absorption on electron target}},''
	\href{http://dx.doi.org/10.1007/JHEP05(2022)191}{{\em JHEP} {\bfseries 05}
		(2022) 191}, [\href{http://arxiv.org/abs/2201.11497}{{\ttfamily
			arXiv:2201.11497}} [hep-ph]].
	
	\bibitem{Dror:2020czw}
	J.~A. Dror, G.~Elor, R.~McGehee, and T.-T. Yu, ``{\it {Absorption of sub-MeV
			fermionic dark matter by electron targets}},''
	\href{http://dx.doi.org/10.1103/PhysRevD.103.035001}{{\em Phys. Rev. D}
		{\bfseries 103} no.~3, (2021) 035001},
	[\href{http://arxiv.org/abs/2011.01940}{{\ttfamily arXiv:2011.01940}}
	[hep-ph]]. [Erratum: Phys.Rev.D 105, 119903 (2022)].
	
	\bibitem{PandaX:2022ood}
	{\bfseries PandaX} Collaboration, D.~Zhang {\em et~al.}, ``{\it {Search for
			Light Fermionic Dark Matter Absorption on Electrons in PandaX-4T}},''
	\href{http://dx.doi.org/10.1103/PhysRevLett.129.161804}{{\em Phys. Rev.
			Lett.} {\bfseries 129} no.~16, (2022) 161804},
	[\href{http://arxiv.org/abs/2206.02339}{{\ttfamily arXiv:2206.02339}}
	[hep-ex]].
	
	\bibitem{PandaX:2024kjp}
	{\bfseries PandaX} Collaboration, T.~Li {\em et~al.}, ``{\it {Searching for
			MeV-scale Axion-like Particles and Dark Photons with PandaX-4T}},''
	[\href{http://arxiv.org/abs/2409.00773}{{\ttfamily arXiv:2409.00773}}
	[hep-ex]].
	
	\bibitem{LZ:2021xov}
	{\bfseries LZ} Collaboration, D.~S. Akerib {\em et~al.}, ``{\it {Projected
			sensitivities of the LUX-ZEPLIN experiment to new physics via low-energy
			electron recoils}},''
	\href{http://dx.doi.org/10.1103/PhysRevD.104.092009}{{\em Phys. Rev. D}
		{\bfseries 104} no.~9, (2021) 092009},
	[\href{http://arxiv.org/abs/2102.11740}{{\ttfamily arXiv:2102.11740}}
	[hep-ex]].
	
	\bibitem{LZ:2023poo}
	{\bfseries LZ} Collaboration, J.~Aalbers {\em et~al.}, ``{\it {Search for new
			physics in low-energy electron recoils from the first LZ exposure}},''
	\href{http://dx.doi.org/10.1103/PhysRevD.108.072006}{{\em Phys. Rev. D}
		{\bfseries 108} no.~7, (2023) 072006},
	[\href{http://arxiv.org/abs/2307.15753}{{\ttfamily arXiv:2307.15753}}
	[hep-ex]].
	
	\bibitem{PandaX:2022xqx}
	{\bfseries PandaX} Collaboration, S.~Li {\em et~al.}, ``{\it {Search for Light
			Dark Matter with Ionization Signals in the PandaX-4T Experiment}},''
	\href{http://dx.doi.org/10.1103/PhysRevLett.130.261001}{{\em Phys. Rev.
			Lett.} {\bfseries 130} no.~26, (2023) 261001},
	[\href{http://arxiv.org/abs/2212.10067}{{\ttfamily arXiv:2212.10067}}
	[hep-ex]].
	
	\bibitem{PandaX:2024muv}
	{\bfseries PandaX} Collaboration, Z.~Bo {\em et~al.}, ``{\it {First Indication
			of Solar $^8$B Neutrino Flux through Coherent Elastic Neutrino-Nucleus
			Scattering in PandaX-4T}},''
	[\href{http://arxiv.org/abs/2407.10892}{{\ttfamily arXiv:2407.10892}}
	[hep-ex]].
	
	\bibitem{Wang:2023wrr}
	X.~Wang, Z.~Lei, Y.~Ju, J.~Liu, N.~Zhou, Y.~Chen, Z.~Wang, X.~Cui, Y.~Meng, and
	L.~Zhao, ``{\it {Design, construction and commissioning of the PandaX-30T
			liquid xenon management system}},''
	\href{http://dx.doi.org/10.1088/1748-0221/18/05/P05028}{{\em JINST}
		{\bfseries 18} no.~05, (2023) P05028},
	[\href{http://arxiv.org/abs/2301.06044}{{\ttfamily arXiv:2301.06044}}
	[physics.ins-det]].
	
	\bibitem{XLZD:2024gxx}
	{\bfseries XLZD} Collaboration, J.~Aalbers {\em et~al.}, ``{\it {The XLZD
			Design Book: Towards the Next-Generation Liquid Xenon Observatory for Dark
			Matter and Neutrino Physics}},''
	[\href{http://arxiv.org/abs/2410.17137}{{\ttfamily arXiv:2410.17137}}
	[hep-ex]].
	
	\bibitem{Dimopoulos:1981au}
	S.~Dimopoulos and S.~Raby, ``{\it {Supercolor}},''
	\href{http://dx.doi.org/10.1016/0550-3213(81)90430-2}{{\em Nucl. Phys. B}
		{\bfseries 192} (1981) 353--368}.
	
	\bibitem{Dimopoulos:1981zb}
	S.~Dimopoulos and H.~Georgi, ``{\it {Softly Broken Supersymmetry and SU(5)}},''
	\href{http://dx.doi.org/10.1016/0550-3213(81)90522-8}{{\em Nucl. Phys. B}
		{\bfseries 193} (1981) 150--162}.
	
	\bibitem{Martin:1997ns}
	S.~P. Martin, ``{\it {A Supersymmetry primer}},''
	\href{http://dx.doi.org/10.1142/9789812839657_0001}{{\em Adv. Ser. Direct.
			High Energy Phys.} {\bfseries 18} (1998) 1--98},
	[\href{http://arxiv.org/abs/hep-ph/9709356}{{\ttfamily
			arXiv:hep-ph/9709356}}].
	
	\bibitem{Kamionkowski:1992mf}
	M.~Kamionkowski and J.~March-Russell, ``{\it {Planck scale physics and the
			Peccei-Quinn mechanism}},''
	\href{http://dx.doi.org/10.1016/0370-2693(92)90492-M}{{\em Phys. Lett. B}
		{\bfseries 282} (1992) 137--141},
	[\href{http://arxiv.org/abs/hep-th/9202003}{{\ttfamily
			arXiv:hep-th/9202003}}].
	
	\bibitem{Holman:1992us}
	R.~Holman, S.~D.~H. Hsu, T.~W. Kephart, E.~W. Kolb, R.~Watkins, and L.~M.
	Widrow, ``{\it {Solutions to the strong CP problem in a world with
			gravity}},'' \href{http://dx.doi.org/10.1016/0370-2693(92)90491-L}{{\em Phys.
			Lett. B} {\bfseries 282} (1992) 132--136},
	[\href{http://arxiv.org/abs/hep-ph/9203206}{{\ttfamily
			arXiv:hep-ph/9203206}}].
	
	\bibitem{Nelson:1983zb}
	A.~E. Nelson, ``{\it {Naturally Weak CP Violation}},''
	\href{http://dx.doi.org/10.1016/0370-2693(84)92025-2}{{\em Phys. Lett. B}
		{\bfseries 136} (1984) 387--391}.
	
	\bibitem{Barr:1984qx}
	S.~M. Barr, ``{\it {Solving the Strong CP Problem Without the Peccei-Quinn
			Symmetry}},'' \href{http://dx.doi.org/10.1103/PhysRevLett.53.329}{{\em Phys.
			Rev. Lett.} {\bfseries 53} (1984) 329}.
	
	\bibitem{Nelson:1984hg}
	A.~E. Nelson, ``{\it {Calculation of $\theta$ Barr}},''
	\href{http://dx.doi.org/10.1016/0370-2693(84)90827-X}{{\em Phys. Lett. B}
		{\bfseries 143} (1984) 165--170}.
	
	\bibitem{Choi:2019omm}
	G.~Choi and T.~T. Yanagida, ``{\it {Solving the strong CP problem with
			horizontal gauge symmetry}},''
	\href{http://dx.doi.org/10.1103/PhysRevD.100.095023}{{\em Phys. Rev. D}
		{\bfseries 100} no.~9, (2019) 095023},
	[\href{http://arxiv.org/abs/1909.04317}{{\ttfamily arXiv:1909.04317}}
	[hep-ph]].
	
	\bibitem{Feruglio:2023uof}
	F.~Feruglio, A.~Strumia, and A.~Titov, ``{\it {Modular invariance and the QCD
			angle}},'' \href{http://dx.doi.org/10.1007/JHEP07(2023)027}{{\em JHEP}
		{\bfseries 07} (2023) 027},
	[\href{http://arxiv.org/abs/2305.08908}{{\ttfamily arXiv:2305.08908}}
	[hep-ph]].
	
	\bibitem{Feruglio:2024ytl}
	F.~Feruglio, M.~Parriciatu, A.~Strumia, and A.~Titov, ``{\it {Solving the
			strong CP problem without axions}},''
	\href{http://dx.doi.org/10.1007/JHEP08(2024)214}{{\em JHEP} {\bfseries 08}
		(2024) 214}, [\href{http://arxiv.org/abs/2406.01689}{{\ttfamily
			arXiv:2406.01689}} [hep-ph]].
	
	\bibitem{Liang:2024wbb}
	Q.~Liang, R.~Okabe, and T.~T. Yanagida, ``{\it {Three-zero texture of
			quark-mass matrices as a solution to the strong CP problem}},''
	[\href{http://arxiv.org/abs/2408.12146}{{\ttfamily arXiv:2408.12146}}
	[hep-ph]].
	
	\bibitem{Cheng:2024yrn}
	Y.~Cheng, J.~Sheng, and T.~T. Yanagida, ``{\it {Detecting a Fifth-Force Gauge
			Boson via Superconducting Josephson Junctions}},''
	[\href{http://arxiv.org/abs/2402.14514}{{\ttfamily arXiv:2402.14514}}
	[hep-ph]].
	
	\bibitem{Cheng:2024zde}
	Y.~Cheng, J.~Lin, J.~Sheng, and T.~T. Yanagida, ``{\it {Proposal for a Quantum
			Mechanical Test of Gravity at Millimeter Scale}},''
	[\href{http://arxiv.org/abs/2405.16222}{{\ttfamily arXiv:2405.16222}}
	[hep-ph]].
	
	\bibitem{Arvanitaki:2017nhi}
	A.~Arvanitaki, S.~Dimopoulos, and K.~Van~Tilburg, ``{\it {Resonant absorption
			of bosonic dark matter in molecules}},''
	\href{http://dx.doi.org/10.1103/PhysRevX.8.041001}{{\em Phys. Rev. X}
		{\bfseries 8} no.~4, (2018) 041001},
	[\href{http://arxiv.org/abs/1709.05354}{{\ttfamily arXiv:1709.05354}}
	[hep-ph]].
	
	\bibitem{Adelberger:2009zz}
	E.~G. Adelberger, J.~H. Gundlach, B.~R. Heckel, S.~Hoedl, and S.~Schlamminger,
	``{\it {Torsion balance experiments: A low-energy frontier of particle
			physics}},'' \href{http://dx.doi.org/10.1016/j.ppnp.2008.08.002}{{\em Prog.
			Part. Nucl. Phys.} {\bfseries 62} (2009) 102--134}.
	
	\bibitem{Wagner:2012ui}
	T.~A. Wagner, S.~Schlamminger, J.~H. Gundlach, and E.~G. Adelberger, ``{\it
		{Torsion-balance tests of the weak equivalence principle}},''
	\href{http://dx.doi.org/10.1088/0264-9381/29/18/184002}{{\em Class. Quant.
			Grav.} {\bfseries 29} (2012) 184002},
	[\href{http://arxiv.org/abs/1207.2442}{{\ttfamily arXiv:1207.2442}} [gr-qc]].
	
	\bibitem{LIGOScientific:2021ffg}
	{\bfseries LIGO Scientific, KAGRA, Virgo} Collaboration, R.~Abbott {\em
		et~al.}, ``{\it {Constraints on dark photon dark matter using data from
			LIGO\textquoteright{}s and Virgo\textquoteright{}s third observing run}},''
	\href{http://dx.doi.org/10.1103/PhysRevD.105.063030}{{\em Phys. Rev. D}
		{\bfseries 105} no.~6, (2022) 063030},
	[\href{http://arxiv.org/abs/2105.13085}{{\ttfamily arXiv:2105.13085}}
	[astro-ph.CO]].
	
	\bibitem{Miller:2023kkd}
	A.~L. Miller and L.~Mendes, ``{\it {First search for ultralight dark matter
			with a space-based gravitational-wave antenna: LISA Pathfinder}},''
	\href{http://dx.doi.org/10.1103/PhysRevD.107.063015}{{\em Phys. Rev. D}
		{\bfseries 107} no.~6, (2023) 063015},
	[\href{http://arxiv.org/abs/2301.08736}{{\ttfamily arXiv:2301.08736}}
	[gr-qc]].
	
	\bibitem{Frerick:2023xnf}
	J.~Frerick, J.~Jaeckel, F.~Kahlhoefer, and K.~Schmidt-Hoberg, ``{\it {Riding
			the dark matter wave: Novel limits on general dark photons from LISA
			Pathfinder}},'' \href{http://dx.doi.org/10.1016/j.physletb.2023.138328}{{\em
			Phys. Lett. B} {\bfseries 848} (2024) 138328},
	[\href{http://arxiv.org/abs/2310.06017}{{\ttfamily arXiv:2310.06017}}
	[hep-ph]].
	
	\bibitem{Fabbrichesi:2020wbt}
	M.~Fabbrichesi, E.~Gabrielli, and G.~Lanfranchi, ``{\it {The Dark Photon}},''
	[\href{http://arxiv.org/abs/2005.01515}{{\ttfamily arXiv:2005.01515}}
	[hep-ph]].
	
\end{thebibliography}
\end{document}